\documentclass[fleqn,usenatbib]{mnras}

\usepackage{newtxtext,newtxmath}

\usepackage[T1]{fontenc}
\usepackage{ae,aecompl}

\usepackage{xcolor}

\usepackage{graphicx}	
\usepackage{amsmath}	
\usepackage{amssymb}	
\usepackage{ulem}       

\title[$A_{FUV}$ as a function of stellar mass and redshift]{UV dust attenuation as a function of stellar mass \\ and its evolution with redshift}

\author[Bogdanoska \& Burgarella]{
Jana Bogdanoska,$^{1}$\thanks{E-mail: jana.bogdanoska@lam.fr}
Denis Burgarella,$^{1}$\thanks{E-mail: denis.burgarella@lam.fr}
\\
$^{1}$Aix-Marseille Universit\'{e}, CNRS, LAM (Laboratoire d'Astrophysique de Marseille) UMR 7326, 13388, France\\
}

\date{Accepted XXX. Received YYY; in original form ZZZ}

\pubyear{20XY}

\begin{document}
\label{firstpage}
\pagerange{\pageref{firstpage}--\pageref{lastpage}}
\maketitle

\begin{abstract}
Studying the UV dust attenuation, as well as its relation to other galaxy parameters such as the stellar mass, plays an important role in multi-wavelength research. This work relates the dust attenuation to the stellar mass of star forming galaxies, and its evolution with redshift. A sample of galaxies with an estimate of the dust attenuation computed from the infrared excess was used. The dust attenuation vs. stellar mass data, separated in redshift bins, was modelled by a single parameter linear function, assuming a nonzero constant apparent dust attenuation for low mass galaxies. But the origin of this effect is still to be determined and several possibilities are explored (actual high dust content, variation of the dust-to-metal ratio, variation of the stars-dust geometry). The best-fitting parameter of this model is then used to study the redshift evolution of the cosmic dust attenuation and is found to be in agreement with results from the literature. This work also gives evidence to a redshift evolution of the dust attenuation - stellar mass relationship, as is suggested by recent works in the highest redshift range.
\end{abstract}

\begin{keywords}
Galaxies: ISM -- (ISM:) dust, extinction -- Infrared: galaxies
\end{keywords}



\section{Introduction}\label{sect:intro}

Galaxies are complex systems containing stars, gas, dust and dark matter, with all of their components interacting with each other to produce a combined multi-wavelength emission: the Spectral Energy Distribution (SED). The SED is the result of the combined emission from each of these components, but it is also influenced by their position relative to each other in space, what we usually refer to as the geometry of the galaxy. In galaxies, luminous stars emit most of the ultraviolet (UV) and optical light, whilst dust influences the light we receive via the process of attenuation. Part of the UV+optical light is absorbed by dust grains and re-emitted in the infrared (IR). So, it is of utmost importance to understand the effects dust has on the multi-wavelength emission of galaxies. The contribution of dust needs to be accounted for in any observations of galaxies if we are to perform a complete census of their components and the physical processes acting on these components.

Interstellar dust is created from the material that is ejected from stars or directly in the interstellar medium (ISM). Dust is built from heavy elements and compounds, such as silicates, carbonaceous materials, silicon carbides, carbonates, etc. \citep{draine_interstellar_2003}. {
It is especially interesting to know how the quantity of dust has evolved throughout cosmic time \citep{takeuchi_model_2005, cucciati_star_2012, burgarella_herschel_2013, madau_cosmic_2014}. All these works agree on a general behavior, that the average Cosmic dust attenuation in galaxies increases from $z=0$ to $z\sim1.5$. This rise is followed by a decrease to $z\sim4$ when only IR data are used. Up to now, this decrease could not be constrained by IR data at higher redshifts, but ALMA and other ground-based millimetre (mm) data now provide further constraints as we try to do in this work.}

Quantifying the amount of dust is challenging. The best method available today is analysing the IR SEDs of galaxies. Most of the light emitted in the far-IR part of the SED is due to the thermal emission of dust \citep[see, e.g., ][]{draine_infrared_2007}. 

However, estimating the amount of light absorbed by dust can also be achieved without far-IR data by using alternate proxy methods. Probably the most popular one is the so-called $\beta$-slope (defined as $f_\lambda \propto \lambda^\beta$ \citep{calzetti_dust_1994}) method  \citep[e.g.][]{bouwens_uv-continuum_2012, bouwens_alma_2016} proposed by  \citet{meurer_dust_1999}, from which the IR Excess (IRX, Eq. \eqref{eq:IRX_def}) can be estimated. This is extremely useful when no IR data is available, which is often the case above $z \approx 4$. However, this relation has been mainly determined until redshift $z\approx 3$ for UV-dominated galaxies \citep[][and references therein]{bowler_obscured_2018} and some departures are observed at higher redshift and for more IR-bright galaxies \citep[e.g.][]{casey_are_2014}. Other common methods for calculating the dust attenuation include using the Balmer decrement (the ratio of the H$_\alpha$ line to the H$_\beta$ line), as well as the H$_\alpha$ line, for which some assumptions need to be made (see, for example, the introduction of the paper by \citet{hao_dust-corrected_2011}). 

The relation between IRX (or $A_{FUV}$) and stellar mass ($M_*$) is yet another
tool that allows to estimate the dust attenuation in galaxies from the stellar mass of galaxies. Because $M_*$ mirrors the previous star formation activity of galaxies, which in turn is responsible for producing dust particles, the stellar mass may be a good, and easy to estimate, tracer of the dust content
in galaxies.

The relationship between the stellar mass and attenuation has been the focus of numerous studies as early as \citep[e.g.][and references therein]{xu_ir_2007, martin_star_2007,buat_infrared_2009}. Most works seem to suggest that there is a linear relation between IRX  and stellar mass over quite a large mass range ($9 \le log M_* \le 12$, \citep[][etc.]{heinis_hermes_2014, pannella_goods-herschel_2015, alvarez-marquez_dust_2016} that is dubbed the  `consensus $z\sim 2-3$ relationship' by \cite{bouwens_alma_2016}. However, this same paper \cite{bouwens_alma_2016} suggests either an evolution of the dust temperature or an evolution of the relationship at large redshift. The latter might be confirmed in the most recent works. For instance,  \citet[E.g.][]{fudamoto_dust_2017, fudamoto_alpine-alma_2020} observe an evolution of the IRX – $M_{star}$ relationship between $z\sim3$ and $z\sim6$ by about 0.24 dex.

This paper is organised as follows. Firstly, we present the possible biases that might affect our analysis in Sect. \ref{sect:biases}. After this, the data we have used to obtain our results are presented in Sect. \ref{sect:data_used}. In Sect. \ref{sect:A_FUV_M_star_main} we present a detailed explanation of the methods implemented in this project and the main results obtained, and we present our findings concerning the relationship of the dust attenuation with stellar mass $A_{FUV}-M_*$ (Sect. \ref{sect:Afuv_Mstar_evol}), and then we use these findings to study the dust attenuation as a function of redshift $A_{FUV}(z)$ (Sect. \ref{sect:A_FUV_z}). In Sect. \ref{sect:cosmic_Afuv} we discuss our estimate of the cosmic dust attenuation and compare it to the values found in the literature. The possible implications of our results are discussed in Sect. \ref{sect:discussion}. We use a \cite{salpeter_luminosity_1955} Initial Mass Function (IMF) for the stellar masses used throughout our work, as well as a $\mathrm{\Lambda CDM}$ cosmology with $(H_0, \Omega_m, \Omega_\Lambda) = (70, 0.3, 0.7)$, where $H_0$ has the units of $\mathrm{km\, s^{-1}\, Mpc^{-1}}$.

\section{Biases introduced due to the nature of observations}\label{sect:biases}

For this work, we compile a lot of data over quite large redshift and stellar mass ranges. We are aware that our approach is not fully complete as the information we can collect on the IR emission of galaxies at all redshifts can hardly be exhaustive and we follow the statistical approaches presented in several papers. The relative performance in UV+optical+near-IR is more favourable than that in far-IR+sub-millimeter. Thus, the detection limits are better for UV-dominated galaxies than for IR-dominated ones. We estimate that we have the following biases:
\begin{enumerate}
    \item[i)]In terms of the redshift:
    \begin{itemize}
       \item At high redshift (z > 2 - 3), a significant part of the less massive galaxies are not detected at all whilst in the far-IR only the dustier objects can be detected. So, it is likely that our high-redshift samples will be biased against low-IRX objects. This would mean that our high-redshift trend might only be seen as upper limits. 
       \item  In the local universe, we are limited by the studied volume that is likely biased against rare objects; IR-bright galaxies are rarer than UV-bright galaxies. But given that both galaxy types are rare, we assume that this bias should not dramatically impact our results.
    \end{itemize}
    \vspace{\baselineskip}
    
     \item[ii)] Concerning the stellar mass:
   \begin{itemize}
       \item Similarly, very massive galaxies are rare and this makes it hard to study their properties. But, like for the previous point, our results should not be strongly impacted. This is confirmed by attempting to modify the upper stellar mass cut-off without changing the global characteristics.
       \item On the contrary, low-mass galaxies are numerous, and usually very faint, so the completeness decreases. It was usually thought that low-mass galaxies suffer from a very low dust attenuation or might even be dust-free. But recent results seem to suggest that this could be too fast of a conclusion: dusty galaxies might appear smaller than they are in reality in the UV because the IR part of the SED is not detected \citep{takeuchi_model_2005, whitaker_constant_2017, alvarez-marquez_rest-frame_2019}. This effect has an important impact on the global redshift evolution of the average dust attenuation because we expect a large number of such objects. 
       
    \end{itemize}
\end{enumerate}

In conclusion, we understand that this work is probably not the final word about the redshift evolution of the IRX-M$_*$ relation. But we need to move beyond the simple linear and constant view about this relationship that was consensual, as more and more results at low and high redshifts suggest that this is not true. The simple fact that this IRX-M$_*$ needs to produce results consistent with the redshift evolution of the average galaxy attenuation presented in the literature  \cite[][etc.]{cucciati_star_2012, burgarella_herschel_2013,madau_cosmic_2014} means that we have to understand it better. One place to start is to use an approach checking that the assumptions are in agreement with the evolution of the dust evolution at cosmological scales.

\section{The data used}\label{sect:data_used}

We select data from the literature to build our final sample. The selection criteria are that the IRX values have been estimated either from a direct IR-to-UV ratio or by SED fitting. We do not keep samples where IRX is estimated from the UV slope $\beta$. Although we think $\beta$ could be a useful dust tracer, it has problems for dusty galaxies that are known not to follow the \cite{meurer_dust_1999} relation \citep{burgarella_star_2005, casey_are_2014}. The reason for this departure is not studied here. We do not use surveys using the Balmer decrements because galaxies selected from emission lines are generally younger and this might impact on our statistics. Since, IRX is the ratio of L$_{IR}$ to  L$_ {UV}$, only the galaxies with measured L$_{IR}$ and L$_ {UV}$ are usable for our study. We also assume that the IRX estimated from SED fitting is close to L$_{IR}$ / L$_ {UV}$ \citep[][for instance]{malek_help_2018}. More precisely, we use the following definition, with the IR luminosity, $L_{{IR}}$, being the total integrated luminosity in the IR, and the UV luminosity, $L_{{FUV}}$, derived from flux measured with a filter, such as, for e.g. \textit{GALEX}.

\begin{align}\label{eq:IRX_def}
\mathrm{IRX} = \log \Bigg( \frac{L_{  {IR}}}{L_{{FUV}}} \Bigg)
\end{align}

In this paper, we call `dust attenuation' the amount of UV energy reprocessed by dust grains, i.e., the net effect caused by the dust grains distributed within the galaxy in a complex geometry. The UV dust attenuation, $A_{FUV}$ is a quantity that tells us by how much the light from the galaxies has been obscured by dust. Practically, for energy-balance based SED fitting, $A_{FUV}$ is the parameter that contains the information of how much of the UV flux has been ``converted'' to IR radiation. We introduce a relationship between IRX and $A_{FUV}$. In this work we use the parametrisation suggested by \citet{hao_dust-corrected_2011}, which has the following form: 

\begin{align}\label{eq:hao_convert}
A_{FUV} = 2.5 \log \big(1 + a_{FUV} \times 10^{\mathrm{IRX}} \big)
\end{align}
with the calibration $a_{FUV} = 0.46 \pm 0.12$. The difference between this conversion and other similar ones (e.g. \cite{buat_dust_2005}) is negligible, this one having the advantage of avoiding giving un-physical negative values for the dust attenuation. 

The data included in this work contains the galaxies from the {GALEX–SDSS–WISE Legacy Catalog (GSWLC) \citep{salim_galex-sdss-wise_2016, salim_dust_2018},} the GOODS-N field \citep{pannella_goods-herschel_2015}, the Cosmic Evolution Survey (COSMOS) field \citep{alvarez-marquez_dust_2016}, as well as the COSMOS field combined with data from the \textit{Herschel} Multi-Tiered Extragalactic Survey (HerMES) program and the Visible and Infrared Survey Telescope for Astronomy (VISTA) \citep{heinis_hermes_2014}, the \textit{Hubble} Ultra Deep Field (HUDF) \citep{bouwens_alma_2016}, as well as some other high-redshift sources \citep{schaerer_new_2015, burgarella_observational_2020, fudamoto_alpine-alma_2020}. A summary of the publications used in this work is given in Table \ref{table:data_sources}. 

The samples in these surveys have been selected by using different criteria. The GSWLC is based on the Main Galaxy Sample (MGS) of the SDSS, and it is magnitude limited. We only include the objects which belong to the MGS, and with SDSS photometry, and with UV data from GALEX. Additionally, we include only the star forming galaxies, as defined by the $SFR - M_* - Z$ relation proposed by \cite{speagle_highly_2014} (their equation 28), and assume a dispersion of 0.3$\,$dex around this relation according to \cite{peng_mass_2010}, outside of which all galaxies are excluded. The different IMFs chosen by the different authors have been taken into account.

\cite{heinis_hermes_2014} use a UV selected sample from the COSMOS field for the three different redshifts presented in their work. Furthermore, \cite{alvarez-marquez_dust_2016} use Lyman-Break Galaxies (LBGs) from the COSMOS field, selected by the classical U-dropout method at redshift $z \approx 3$. The HUDF studied by \cite{bouwens_alma_2016} also contains a UV-selected sample of LBGs. \cite{fudamoto_dust_2017} include UV-selected massive star-forming galaxies from the COSMOS field.\cite{fudamoto_alpine-alma_2020} use the ALPINE sample \citep{lefevre_alpine-alma_2019, bethermin_alpine-alma_2020, faisst_alpine-alma_2020}. {The GSWLC contains nearby galaxies. For most of them, the spectra have been measured. They have redshifts $z<0.3$, and a mean redshift of $z=0.1$.} The other authors use various techniques to determine the redshifts of the galaxies within their sample; \citet{pannella_goods-herschel_2015},  and \citet{bouwens_uv-continuum_2012} use the software \textsc{eazy}, whilst \citet{heinis_hermes_2014} and \citet{alvarez-marquez_dust_2016} use an i-band selected COSMOS catalogue  produced with the software \textsc{LePhare} \citep{ilbert_cosmos_2009}. For the objects studied by \citet{schaerer_new_2015}, \cite{burgarella_observational_2020}, and \cite{fudamoto_alpine-alma_2020} the redshifts have been spectroscopically determined.

The stellar masses of the galaxies included in our final sample have been calculated by similar, but not identical methods by the different groups. Most of the authors have used the method of SED fitting \citep[e.g.][]{walcher_fitting_2011}, by assuming a different IMF. \citet{pannella_goods-herschel_2015}, and \citet{bouwens_alma_2016} use a \citet{salpeter_luminosity_1955} IMF, while {\cite{salim_galex-sdss-wise_2016, salim_dust_2018},} \citet{heinis_hermes_2014}, \cite{alvarez-marquez_dust_2016}, \citet{fudamoto_dust_2017}, \cite{burgarella_observational_2020}, and \cite{fudamoto_alpine-alma_2020} use a \citet{chabrier_galactic_2003} IMF. They all implement the \citet{bruzual_stellar_2003} single stellar populations, as well as an exponentially declining Star Formation History (SFH), {except \cite{salim_galex-sdss-wise_2016, salim_dust_2018}, who use a two-component exponential SFH, and \cite{fudamoto_alpine-alma_2020} who use a constant SFH} \citep{faisst_alpine-alma_2020}. Some of the authors do test other SFHs in their work, concluding that its impact is negligible on the results. \citet{schaerer_new_2015} use a different calibration for the stellar masses, obtained by the same authors in another work (cited as in prep. in \citet{schaerer_new_2015} and private communication).  
 
 To account for the different IMF used, a correction has been applied so that all of the data matches a \citet{salpeter_luminosity_1955} IMF. The conversion from \citet{chabrier_galactic_2003} to \citet{salpeter_luminosity_1955} IMF is a multiplicative factor in terms of the mass, or an additive constant when the mass is presented in logarithmic units. The correction we apply is the one given in Eq. 12 by \citet{longhetti_stellar_2009}:

\begin{align}\label{eq:logM_conversion}
   \log M_{*[Salpeter]} =  \log M_{*[Chabrier]} + 0.26 \, {dex} 
\end{align}

\begin{table*}
\centering
\caption{Summary of the literature used to obtain the data, and values of the redshift bins from each reference. For \citet{salim_galex-sdss-wise_2016, salim_dust_2018} the redshift is taken to be $z=0.1$, which is the mean value of the redshifts of all the galaxies in the sample. A range is given for the \citet{bouwens_alma_2016} because individual galaxies are used, and the separation of the bins is performed specifically for this work.}
\label{table:data_sources}
\begin{tabular}{l c c }
\hline\hline
{Reference}	&	{Galaxy Count}	&	{z}	\\ \hline
\citet{salim_galex-sdss-wise_2016, salim_dust_2018}		&	$\approx 400,000$				&	$ < 0.3$	\\ 
\citet{pannella_goods-herschel_2015}	&	$\approx 50,000$, stacked	&	$0.7, 1, 1.3, 1.7, 2.3, 3.3$	\\ 
\citet{heinis_hermes_2014}	&	$\approx 40,000$, stacked	&	$1.5, 3, 4$	\\ 
\citet{alvarez-marquez_dust_2016} &	$\approx 22,000$, stacked	&	$3$	\\
\cite{fudamoto_dust_2017}   &	39					&	$3.2$	\\
\citet{schaerer_new_2015}	&	5  	&	$6.5 - 7.5$	\\ 
\citet{fudamoto_alpine-alma_2020}	&	23					&	$4-5$	\\
\citet{bouwens_alma_2016}	&	78					&	$4-10$	\\
\citet{burgarella_observational_2020}	&	18					&	$5-10$	\\

\hline
\end{tabular}
\end{table*}

Two different types of data are included in this work: data of individual galaxies \citep{salim_galex-sdss-wise_2016, salim_dust_2018, schaerer_new_2015,  fudamoto_dust_2017, burgarella_observational_2020, fudamoto_alpine-alma_2020}, and stacked data \citep{pannella_goods-herschel_2015, heinis_hermes_2014, alvarez-marquez_dust_2016}. The data in the paper of \cite{bouwens_alma_2016} has been stacked, but in this work, we use the photometric data of the individual galaxies directly and we perform an SED fitting on individual galaxies by using \textsc{cigale} \citep{burgarella_star_2005, noll_analysis_2009, boquien_cigale_2019} (Table  \ref{table:data_sources}).


\section{Evolution of the dust attenuation}\label{sect:A_FUV_M_star_main}

 \begin{figure*}
\centering
   \includegraphics[width = \textwidth]{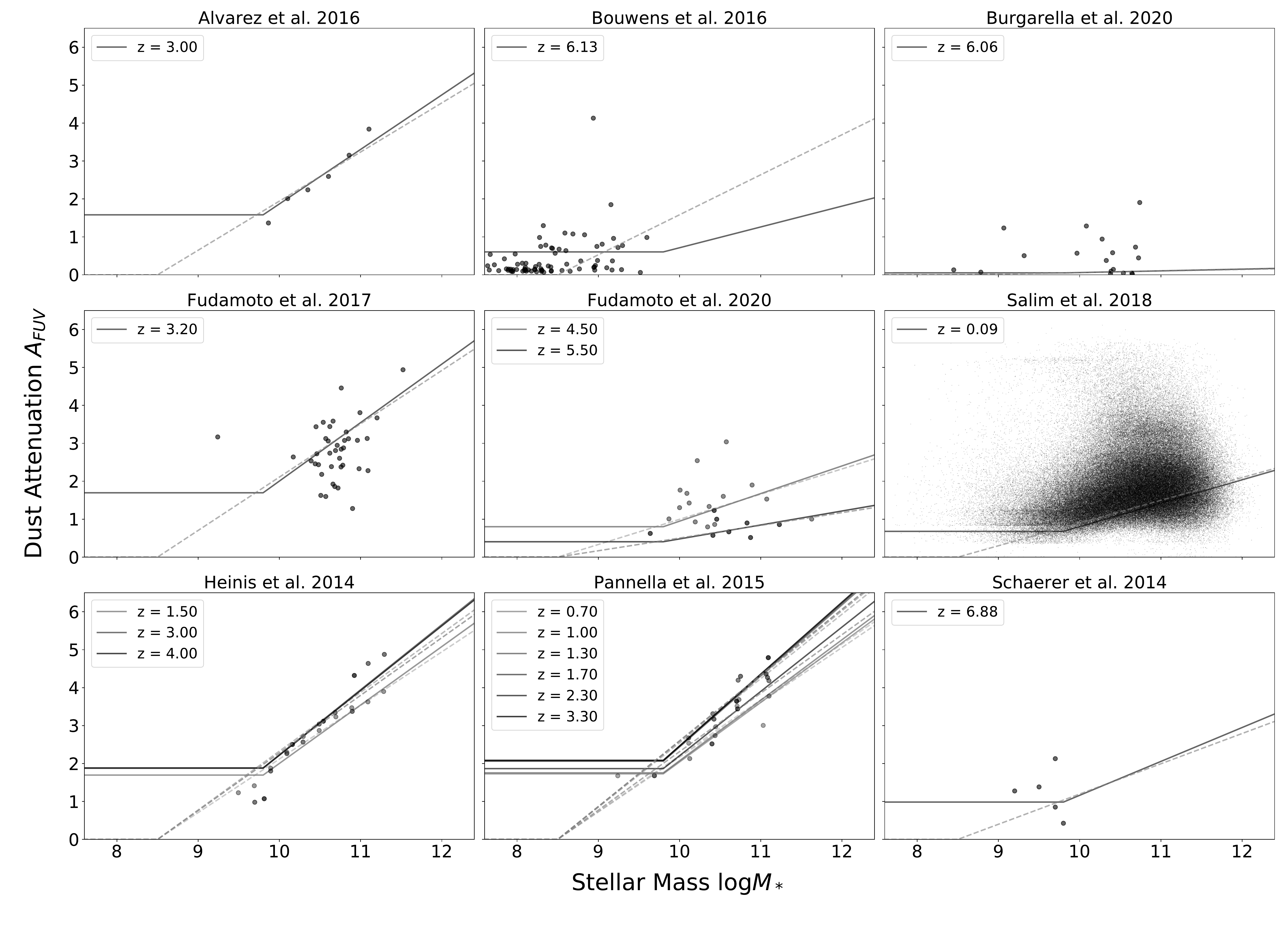}
     \caption{
     The dependence of the UV dust attenuation on stellar mass, showing the data from several references, along with the best-fitting model for the same redshift. 
     Each panel represents the data from a different paper, with multiple different lines within one panel are models for different redshift bins. The dashed lines represent the model proposed in Eq. \eqref{eq:Afuv_M*}, while the full lines show the model of Eq. \eqref{eq:Afuv_M*_break}.
     }
     \label{fig:AFUV_Mstar_data}
\end{figure*}

This section is dedicated to studying the relationship between the stellar mass of star forming galaxies and their average dust attenuation in the FUV, as estimated by the IRX of the galaxies' SEDs. We intend to extend our study to the evolution of this relationship with cosmic times, as well as estimate the average (cosmic) dust attenuation.

We start by dividing the data in redshift bins. In the case where the data has been already divided into redshift bins these bins are kept. In the case of different authors using the same redshift, separate bins are assigned. This means that, for example, for redshift $z = 3$ we have two bins, one from \cite{heinis_hermes_2014} and another from \cite{alvarez-marquez_dust_2016}. The data that are given for individual galaxies are divided in two bins for \cite{bouwens_alma_2016} and \cite{burgarella_observational_2020}, and kept as a single bin for \cite{salim_dust_2018} and \cite{schaerer_new_2015}. The data and the best fitting function are given in Fig. \ref{fig:AFUV_Mstar_data}. Each panel shows the data given by the different authors, separated in redshift bins where appropriate. 

We fit the dust attenuation with a function that depends on two parameters, stellar mass and redshift, i.e.  $A_{FUV}(M_*, z)$, and we express it as a product of two independent functions, each of which has only one variable, namely $A_{FUV} = f(M_*) \times a(z)$.

\subsection{Evolution of the dust attenuation with stellar mass} \label{sect:Afuv_Mstar_evol}

The stellar mass dependence $f(M_*)$ has been studied before \citep[e.g.][]{heinis_hermes_2014, pannella_goods-herschel_2015, alvarez-marquez_dust_2016, bouwens_alma_2016, mclure_dust_2018}, and is usually assumed to be linear either in IRX or directly in $A_{FUV}$. Because we adopt a more global approach, we eventually modify this dependence and assume a broken line by using a function that is linear until a certain value for the stellar mass, and constant below. The justification for this shape is explained in detail in Sect. \ref{sect:modify_Afuv_M}. The function has the same shape for any redshift, however the scaling factor $a$ is not. This constant affects both the value for the function where it is constant and the slope for the linear part. The function is as follows: 

\begin{align}\label{eq:Afuv_M*}
A_{FUV}(\log M_*) = a (\log M_* - 8.5)
\end{align}

As mentioned before, the fitting of the data, shown in Fig. \ref{fig:AFUV_Mstar_data}, is done by setting only $a$ as a free parameter. However, the other parameter which is the x-intercept of the function, namely the value $8.5$ is kept constant. This value also comes from the fitting of the data; once the best-fitting value for $a$ was found, the $\chi ^2$ of the fit is computed for each redshift bin. For different values of the intercept, the values of the $\chi ^2$ were compared. The final values given in Eq. \eqref{eq:Afuv_M*} are the ones giving the lowest $\chi ^2$ on average between all redshift bins. This is done, opposed to directly fitting both parameters in each redshift bins, in order to keep the redshift dependence only in $a$, and have one single value for the intercept. 

Our goal is to find a function that describes the dependence of $A_{FUV}$ on both stellar mass and redshift, and in this section we explore separately the dependence only on stellar mass. We propose to fit the $A_{FUV} - M_*$ relationship with a linear function, multiplied by a factor $a$. This relatively simple function allows us to advance easily to the relationship $A_{FUV} - z$ presented in Sect. \ref{sect:A_FUV_z}. However, we will explore in the following section  (Sect. \ref{sect:modify_Afuv_M}), the possibility of modifying this function.

\subsection{Evolution of the dust attenuation with  redshift}\label{sect:A_FUV_z}

\begin{figure}
 \resizebox{\hsize}{!}{\includegraphics{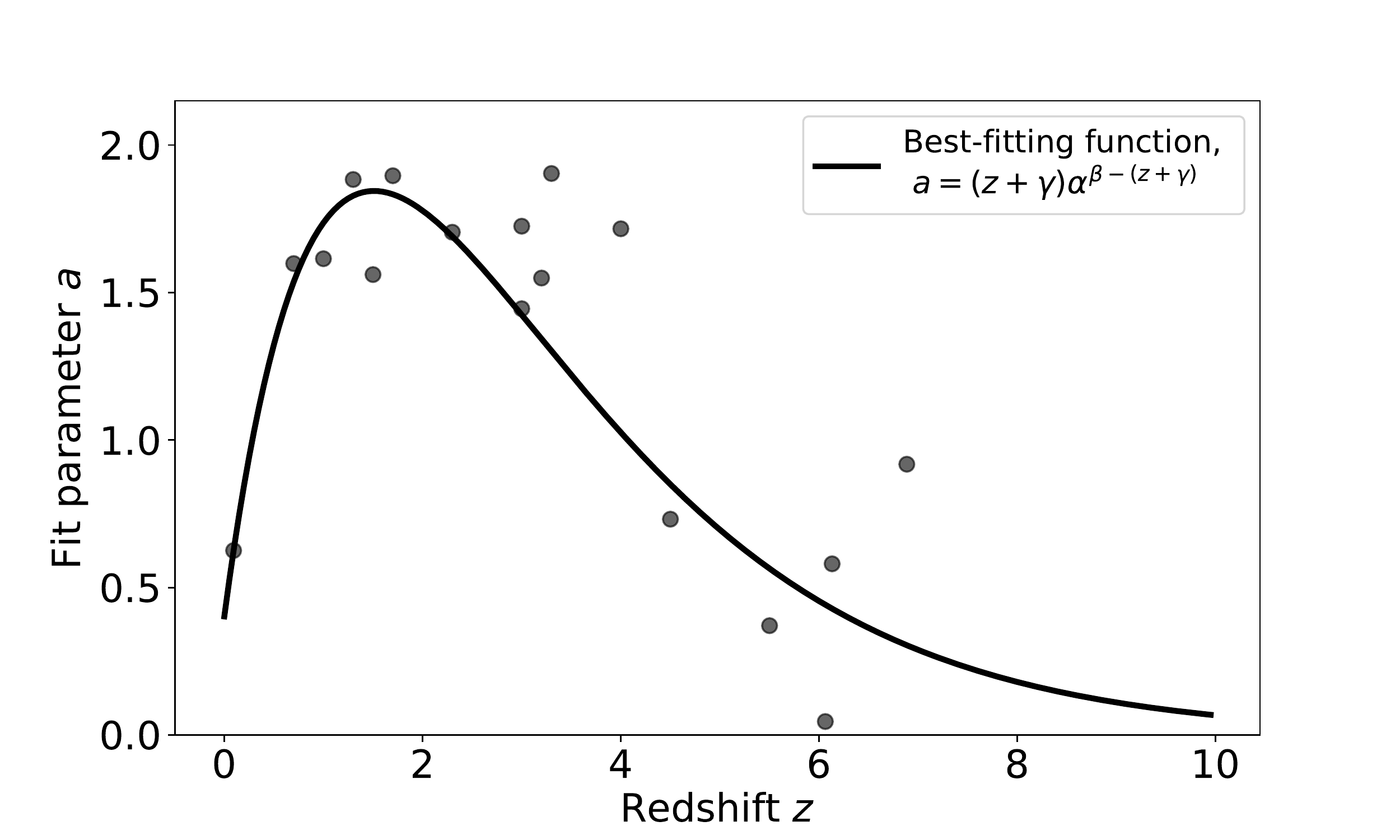}}
 \caption{Fitting the parameter $a$ from the ${A_{FUV}}-M_*$ relationship in each redshift. Each point has been obtained by fitting the available data in that redshift. The black line represents the fit of these points, as fitted with the function of Eq. \eqref{eq:Afuv_z}.}

  \label{fig:a_fit}
\end{figure}

In this section we will study the shape of $a(z)$, which comes from the fitting parameter of the $A_{FUV} - M_*$ relation, with the difference that this time it is not a simple constant $a$, but a function of redshift, i.e. $a(z)$. 

We show in Fig. \ref{fig:a_fit} the evolution in redshift of the parameter $a$, that have been obtained from the best-fitting function of the data, as presented in Sect. \ref{sect:Afuv_Mstar_evol}, and described by Eq. \eqref{eq:Afuv_M*}. Next, we use the points shown in Fig. \ref{fig:a_fit} to find a functional form for $a(z)$, by fitting these coefficients. The black line represented in Fig. \ref{fig:a_fit} is the best fitting function, described as: 

\begin{align}\label{eq:Afuv_z}
a(z) = \left(z+\gamma\right)\cdot \alpha^{\left(\beta-\left(z+\gamma\right)\right)}, 
\end{align} 
the coefficients have the following values: $\alpha = {1.84 \pm 0.12}$, $\beta = {1.59 \pm 0.12}$, and $\gamma = {0.17 \pm 0.04}$. We propose this function as it has a similar shape to the one used by \cite{madau_cosmic_2014} and \cite{burgarella_herschel_2013}, but giving a better fit.

\section{Comparison with the cosmic dust attenuation} \label{sect:cosmic_Afuv}

 \begin{figure}
 \resizebox{\hsize}{!}{\includegraphics{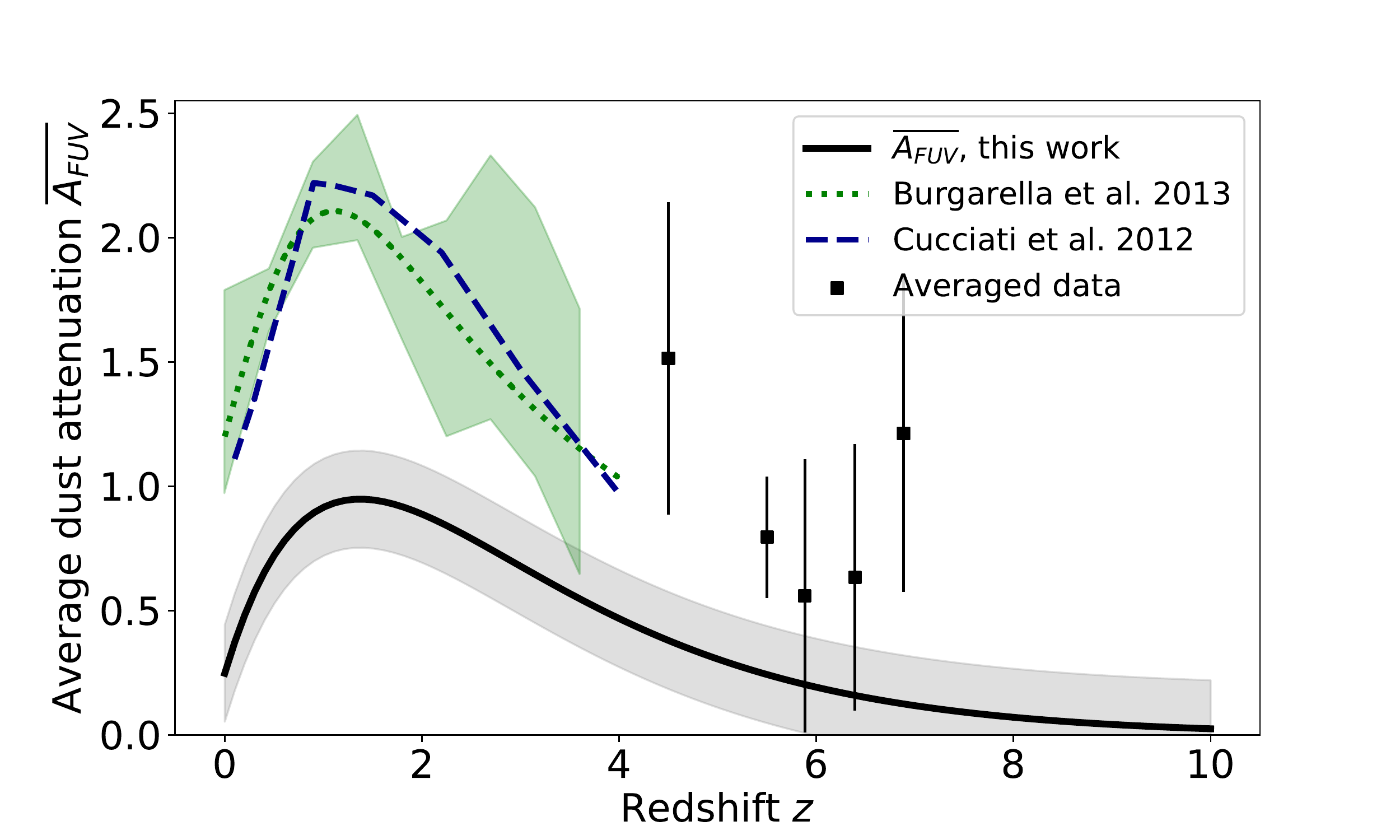}}
 \caption{ 
The evolution of the dust attenuation in the FUV with redshift. The full black line represents the integrated average dust attenuation, calculated using Eq. \eqref{eq:mean_value}, with the model of Eqs. \eqref{eq:Afuv_M*} and \eqref{eq:Afuv_z} (details of the calculations in Appendix \ref{sect:appendix_MF}). Only data with $\log M_* > 9$ has been included in the computations, and consequently $\overline{A_{FUV}}$ has been computed estimating the integrals of Eq. \eqref{eq:mean_value} within the limits of $9 < \log M_* < 14$. 
 The shaded area around the full black line corresponds to the total estimated 1-$\sigma$ uncertainty of the parameters of Eq. \eqref{eq:Afuv_M*}, namely the uncertainty of the intercept of the function estimated with the $\chi ^2$ described in Sect. \ref{sect:Afuv_Mstar_evol}, alongside the errors of the fitting for the coefficients of Eq. \eqref{eq:Afuv_z}. 
The points represent the mean value of the data we included in our work (Fig. \ref{fig:AFUV_Mstar_data}) for $z > 4$, with the errorbars representing the 1-$\sigma$ dispersion around the mean value.  
 The dotted green line and the shaded green area surrounding it come from the work of \citet{burgarella_herschel_2013}, the line is the best-fitting model and the shaded area are the error bars. The dashed dark blue line shows the results of \citet{cucciati_star_2012}. 
 }

     \label{fig:AfuvZ_average}
\end{figure}

 \begin{figure}
 \resizebox{\hsize}{!}{\includegraphics{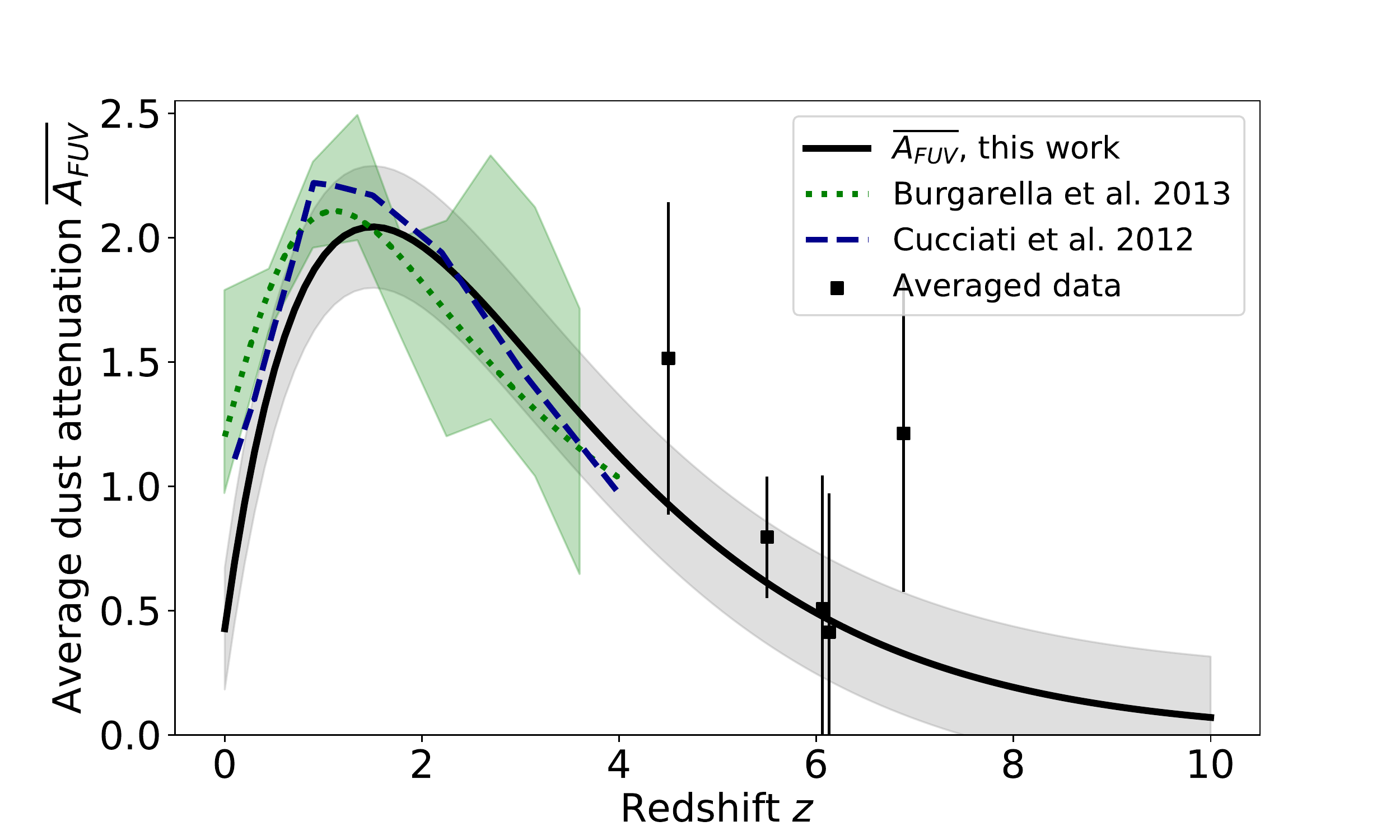}}
 \caption{
The evolution of the dust attenuation in the FUV with redshift. The full black line represents the integrated average dust attenuation, calculated using Eq. \ref{eq:mean_value}, with the model of Eqs. \eqref{eq:Afuv_z} and \eqref{eq:Afuv_M*_break}. In this case $\overline{A_{FUV}}$ has been computed using the limits of $6 < \log M_* < 14$, analogous to the limits of integration used in the work of Burgarella et al. (2013).
The shaded area around the full black line corresponds to the total estimated 1-$\sigma$ uncertainty of the parameters of Eq. \eqref{eq:Afuv_M*_break}, namely the uncertainty of the intercept of the function and the position of the break estimated with the $\chi ^2$ described in Sect. \ref{sect:modify_Afuv_M}, alongside the errors of the fitting for the coefficients of Eq. \eqref{eq:Afuv_z}. We note, however, that the uncertainties might be under-evaluated at low redshift. The origin of this under-evaluation is not clear but maybe it is due to a relatively well fitting of Eq. \eqref{eq:Afuv_z} for $z = 0$.
 Similarly to Fig. \ref{fig:AfuvZ_average}, the points represent the mean value of the used data for $z > 4$, however, they have slightly different values than those of Fig. \ref{fig:AfuvZ_average} due to the objects with $\log M_* < 9$ that were included in computing the mean. 
 The dotted green line, the shaded green area surrounding it, and the dashed dark blue are the same as in Fig. \ref{fig:AfuvZ_average}. 
 }
     \label{fig:AfuvZ_average_break}
\end{figure}

We compare Fig. \ref{fig:a_fit} to the available results in the literature. This work assumes that the $A_{FUV} - M*$ relation is able to represent the dust attenuation of all star forming galaxies given their stellar mass. According to this, if we wish to compute the average dust attenuation, i.e. the cosmic dust attenuation, we need to include \textit{all} of the star forming galaxies. This is why we compute the weighted average of the dust attenuation, the weights being the Mass Function (MF) for star forming galaxies. The details of this computation are given in Appendix \ref{sect:appendix_MF}.

At any redshift data for galaxies with log M* < 9 are scarce. So, in this section we start by using only galaxies with masses in the range $9 < \log M_* < 14$ (Fig. \ref{fig:AfuvZ_average}), and later we include all of the objects (Sect. \ref{sect:modify_Afuv_M}). From Fig. \ref{fig:AfuvZ_average}, is appears that the absolute level  of $A_{FUV}$ as a function of the redshift is too low, as compared to the literature \citep{cucciati_star_2012, burgarella_herschel_2013, madau_cosmic_2014}. The reason for this can be found in the way we compute $\overline{A_{FUV}}$, more precisely in the shape of the mass function. Let us take as an example the MF proposed by \cite{wright_gama/g10-cosmos/3d-hst_2018} in their Fig. 1. We can see that for low mass galaxies, such as $\log M_* = 7$ for the first three redshift bins (for $z < 0.2$), the number is as high as $10^{-1}$ galaxies per unit volume, whilst in the higher stellar mass range, for e.g. for $\log M = 11$ the number is two orders of magnitude lower, so it is roughly $10^{-3}$ galaxies per unit volume. Modelling the MF with a Schechter function means that when including galaxies with down to $\log M \sim 6$, our computation will be heavily influenced by this large number of low mass objects. However, by defining the function which describes $A_{FUV} - M_*$ with Eq. \eqref{eq:Afuv_M*}, we have set $A_{FUV}$ to be zero for galaxies with $\log M_* < 8.5$, i.e. where the single-line models gives $A_{FUV} < 0$ we set $A_{FUV} = 0$.

\subsection{Modification of the function for \texorpdfstring{${A_{FUV}}-M*$}{}}\label{sect:modify_Afuv_M}

As it is obvious from Fig. \ref{fig:AfuvZ_average}, there seems to be a problem with fitting the $A_{FUV} - M_*$ relationship with Eq. \eqref{eq:Afuv_M*}. We suspect that the problem lies within the different numbers of galaxies with low stellar mass versus galaxies with higher stellar mass, which dictates that the apparent average cosmic value of the dust attenuation will be shifted towards the values attributed to the low-mass galaxies. Possible physical explanations for this discrepancy will be detailed in the discussion (Sect. \ref{sect:discussion}). But, here we explore the possibility of modifying the $A_{FUV} - M_*$ relation in such a way that when comparing to the literature for the evolution of  $\overline{A_{FUV}}$ with cosmic time, we get results that are of similar values.

The amount of data for low-mass galaxies, especially at higher redshift, is quite low to date. Consequently, determining the shape of the $A_{FUV} - M_*$ function solely from data is virtually impossible in this stellar mass range. We propose to test the simplest possible form as a first approximation: a constant. We wish to keep the function continuous, as well as to choose the parameters presented in Sect. \ref{sect:Afuv_Mstar_evol} that fit the data as well as possible. Thus we modify Eq. \eqref{eq:Afuv_M*} to have the following form:

\begin{align}\label{eq:Afuv_M*_break}
A_{FUV}(\log M_*) = a \begin{cases} 
1.1 , &\log M_* <  9.8\\ 
\log M_* - 8.7 , &\log M_* \geq 9.8
\end{cases}
\end{align}

This function now has two parameters that determine its shape: the break where the function changes from constant to linear (in Eq. \eqref{eq:Afuv_M*_break} this parameter is equal to $\log M_* = 9.8 \pm 0.1$) and the x-intercept of the linear part (equal to $8.7 \pm 0.1$ in Eq. \eqref{eq:Afuv_M*_break}, and equal to $8.5$ in Eq. \eqref{eq:Afuv_M*}, also expressed in units of $\log M_*$). The value of the constant part is the value of the linear part computed for $\log M_* = 9.8$, to ensure continuity of the function. These two parameters, the break and the intercept, were determined by finding the lowest $\chi^2$ produced by both parameters simultaneously.

The coefficient $a$ is the same as described in Sect. \ref{sect:A_FUV_z}. The redshift dependence of the cosmic dust attenuation, computed using Eq. \eqref{eq:Afuv_M*_break} is presented in Fig. \ref{fig:AfuvZ_average_break}. It should be noted that replacing the function of Eq. \eqref{eq:Afuv_M*} with that of Eq. \eqref{eq:Afuv_M*_break}, produces a difference in the best-fitting values of the parameters of Eq. \eqref{eq:Afuv_z}, thus in this case we give the following values: $\alpha = {1.84 \pm 0.11}$, $\beta = {1.84 \pm 0.12}$, and $\gamma = {0.14 \pm 0.04}$.

\section{Discussion}\label{sect:discussion}

Understanding in its entirety the processes of formation and evolution of galaxies requires knowledge on the subject of cosmic dust. This work makes use of the fact that the scientific community is apt at and confident about estimating the stellar mass $M_*$ of a galaxy. We are striving towards developing a model which would be able to estimate the FUV dust attenuation of a galaxy from its stellar mass and redshift. This would, in turn, enable us to better estimate the Star Formation Rate (SFR) of a galaxy, and give us insight into the overall evolution of distant galaxies. 	

Some of the recently published work argues against this approach, as the evolution of the $A_{FUV} - M_*$ relation has been doubted \citep{heinis_hermes_2014, bouwens_alma_2016,whitaker_constant_2017}. One of the goals of this work is to review this question with a larger set of data covering a large redshift range from $z=0$ to the highest redshift galaxies. On the other hand, \cite{bernhard_modelling_2014} suggest that there is some evolution, only limited to $z<1$. They use the relation from \cite{heinis_hermes_2014} as the basis, and for $z<1$ they vary the normalisation as $IRX'_0 = IRX_0 -0.5 \times (1-z)$. We attempted to add this relation to Fig. \ref{fig:AfuvZ_average_break}, but as the \cite{heinis_hermes_2014} relation is only defined for $\log M_* > 9.5 $, we encountered the same problem as using Eq. \eqref{eq:Afuv_M*} instead of Eq. \eqref{eq:Afuv_M*_break}. 
 
 To assess the validity of our work, we compare to the relevant literature \citep{cucciati_star_2012, burgarella_herschel_2013, madau_cosmic_2014} until redshift $z < 4$. We can see in Fig. \ref{fig:AfuvZ_average_break} that the values obtained by our models (Eq. \eqref{eq:Afuv_M*_break}, combined with Eq. \eqref{eq:Afuv_z}) do indeed follow a similar trend as those found in the literature, even with results obtained by using a completely different method to estimate the dust attenuation. The green dotted line shows the best-fitting model proposed by \cite{burgarella_herschel_2013}, with the green shaded area showing their uncertainties. The work of \cite{burgarella_herschel_2013} is based on the study of IR and UV luminosity functions; the IRX in this work is computed as the ratio of the IR luminosity density to the UV luminosity density, estimated at different redshifts. We also compare our results to the work of \cite{cucciati_star_2012}, who use a different method of estimating the same parameter; they estimate the intrinsic colour excess $E(B-V)$ and by using the starburst reddening curve given by \cite{calzetti_dust_2000}, they estimate the dust attenuation in the FUV. Their results are shown by the dashed dark blue line in Fig. \ref{fig:AfuvZ_average_break}.

To compare our work to that of \cite{burgarella_herschel_2013}, we must compute the average value of the dust attenuation using the same stellar mass range, meaning, perform the integration within the same limits, as explained in Appendix \ref{sect:appendix_MF}. \cite{burgarella_herschel_2013} state that the LFs are integrated within the range $\log_{10} (L[L_\odot]) = [7, 14]$. The corresponding range, converted to units of stellar mass by the use of the $\log M_*-M_{UV}$ relation, with $M_{UV}$ the absolute UV magnitude, given by \cite{song_evolution_2016}, is $\log_{10} (M[M_\odot]) = [6, 14]$. \cite{cucciati_star_2012} do not include a mass or luminosity range in their computations, so we cannot compare in an analogous way. 

\subsection{The apparent dust attenuation of low-mass galaxies}\label{sect:low-mass}

Using our method described in Sect. \ref{sect:A_FUV_M_star_main}, we implicitly make the assumption that the shape of the function of the $A_{FUV}-M_*$ relation is the same throughout all redshifts. We can see, for example, in the data from \cite{salim_galex-sdss-wise_2016, salim_dust_2018} (Fig. \ref{fig:AFUV_Mstar_data}), a clear flattening towards the lower mass end. This sample includes galaxies with stellar masses as low as $\log M_* \approx 7$, who have a large dispersion in the $A_{FUV}$, with the values of the dust attenuation being as low as $A_{FUV} = 0.25$, and with some objects having a value as high as $A_{FUV} = 5$. We can conclude from this data that the low-mass galaxies have a large scatter in their FUV dust attenuation and a mean significantly different from zero. We believe this justifies approximating this part of the $A_{FUV}-M_*$ dependence with a nonzero constant average value. 

On the theoretical side, the work of \cite{cousin_g..s.._2019}, where they use the semi-analytical models called G.A.S. predicts the dust attenuation of galaxies by computing the IRX. We can see in their Fig. 9 a flattening for lower stellar masses similar to the one we find using our parametrisation. For reference, $\log \mathrm{IRX}  =  0.25$ corresponds to $A_{FUV}  =  0.97$, according to the relation of \cite{hao_dust-corrected_2011}. Additionally, they show in Fig. 11 an evolution of the IRX-Mass relation.

A flattening and an increased scatter of the $A_{FUV} - M_*$ relation for galaxies with low stellar masses ($\log M_* < 8$) can equally be seen in simulations, such as the high-resolution cosmological zoom-in simulations FIRE-2 \citep{ma_dust_2019}. They compute galaxy SEDs and mock images using a radiative transfer code adopting a Small Magellanic Cloud (SMC)-type dust grain size distribution, which is preferred for galaxies at higher redshift. In \cite{ma_dust_2019}, one of the important parameters is the  dust-to-metal ratio ($M_{dust} = f_{dust} M_{metal}$). In their simulations, $f_{dust}$ includes all the processes in the dust cycle (dust production, growth and destruction), and is taken to be constant for a given model. Multiple values are tested (see their Fig. 14) and it is suggested that $f_{dust}$ could be observationally constrained. They find that for low-mass galaxies there is a larger scatter and possible flattening, regardless of the value of $f_{dust}$ and independently of the redshift. However,
two main parameters (the opacity $\kappa _{dust}$ and $f _{dust}$) can impact on the absorption coefficient $\alpha$
$\propto$ $\kappa _{dust}$ $f _{dust}$ that enters the radiative transfer equation and sets the dust temperature and emissivity that sets a degeneracy. So the unexplained behavior at low stellar mass can have different origins linked to an intrinsic dust attenuation with a surprising large amount of dust in these low-mass galaxies but other origins are possible like a more clumpy geometry where young stars would be included in dense dusty shells, or a more ''bursty'' nature of the star formation or finally a modification of $f _{dust}$ with the metallicity.

On the observational side, we now see more and more evidence that the simple low-mass low-$A_{FUV}$ assumption might not be fully valid.

A scatter is suggested in the Fig. 2 of \cite{whitaker_constant_2017} which shows that for stellar masses around $\log M_*  =  9$, the dust attenuation can be in the range $0.5 < A_{FUV} < 2.5$. This is in agreement with our results (outliers in the top left corner of the same figure; the value of $f_\mathrm{obscured} = 0.55$ corresponds to $A_{FUV}  =  0.5$ and the value of $f_\mathrm{obscured} = 0.95$ corresponds to $A_{FUV}  =  2.5$, after first converting $f_\mathrm{obscured}$ to IRX, and then using the \cite{hao_dust-corrected_2011} relation to get $A_{FUV}$).

We can also notice the objects reported in the work of \cite{takeuchi_star_2010} (their Fig. 16), where we see galaxies with stellar masses as low as $7 < \log M_* < 8$ which have $A_{FUV}$ values in the range $0.3 < A_{FUV} < 4.1$. Indeed, to be completely certain that low mass galaxies have a higher average dust attenuation than is predicted by previous work, we would need more statistics for fainter galaxies. The next steps of this work would include taking into account the scatter around the average value proposed by our model, so instead of proposing one average value for all low-mass galaxies, we could give a range of possible values. This is, however, beyond the scope of this paper.

The behavior of this $A_{FUV} vs. \log M_*$ law at low mass is puzzling but is required if we wish to match both the data from \citep{salim_galex-sdss-wise_2016}, other less complete studies at low redshift cited above but also IZw18 or SBS~0335-052 \citep{remy-ruyer_linking_2015, hunt_new_2003, lebouteiller_star_2019, reines_new_2008, cormier_new_2017, wu_dust_2007, cannon_dust_2002}. This is also required to match the $A_{FUV}(z)$ shape, consistently obtained using a large variety of methods as illustrated in, e.g., \citet{madau_cosmic_2014}.

We do not have any strong explanation, yet, but we could speculate that dust is building very fast in low-mass objects \cite[see, e.g.][]{burgarella_observational_2020} and could quickly reach a minimum (statistical) threshold close to the value found here, qualitatively speaking, because dust builds from metals, in a way similar to the pop.III - pop.II critical metallicity, \citep[see, e.g., ][]{bromm_fragmentation_2001, schneider_first_2002, jaacks_baseline_2018}. If so, we could observe a flattening of the relation between dust attenuation and metallicity. It is very difficult to confirm this but such a flattening of the dust attenuation as a function of the metallicity is not excluded by \citet{garn_predicting_2010, xiao_dust_2012, koyama_predicting_2015} and \citet{qin_universal_2019}. This means that the present relation does not remain infinitely flat but should present a sharp rise at some low stellar mass.

\subsection{The evolution of the \texorpdfstring{${A_{FUV}}-M*$}{} relation with the redshift}\label{sect:redshift-evolution}

At high redshift, we also have more and more evidence from objects extracted from deep ALMA maps that the `consensus' law is not valid anymore. \citet[E.g.][]{fudamoto_dust_2017, fudamoto_alpine-alma_2020} suggest that there is a significant redshift evolution of the IRX – $M_{star}$ relation between $z\sim3$ and $z\sim6$ by about 0.24 dex. This hypothesis is also supported by the rest of the data at $z>4-5$ presented in this paper.
 
In short, the low stellar mass galaxies at low redshift  exhibit a large scatter in $A_{FUV}$, which can be fitted by a constant function. Based on this, we make the assumption that the $A_{FUV}-M_*$ relation is constant in this mass range for all redshifts. We do leave the option of the value of the constant to vary with redshift, through the fitting of the parameter $a$ in Eq. \eqref{eq:Afuv_M*}. Considering that for higher redshifts, we do not have low-mass galaxy data, we attempt to make up for this by assuming that the evolution of the average $A_{FUV}$ follows the function proposed by \cite{burgarella_herschel_2013}. We then attempt to find such a parametrisation for $A_{FUV}-M_*$ that  would give similar values for the $A_{FUV}-z$ relationship to those of \cite{burgarella_herschel_2013}, when weighted by the MF and integrated to compute the mean $A_{FUV}$. So, fitting the literature data (from Table \ref{table:data_sources}) assures we have a function that reproduces the data well in the higher mass range, while comparing to \cite{burgarella_herschel_2013} compensates for the lack of data in the low mass range, and gives us a prediction on which values we could expect for the $A_{FUV}$ of such objects. 

We are interested in gaining as much knowledge about the early Universe as possible, and understanding the dust attenuation far back in cosmic time is no exception. As can be seen in Table \ref{table:data_sources}, we have included some galaxies with high redshift, however, it is only a small number of objects and only until redshift $z < 8$. So, until more observations are carried out and more advanced telescopes are used, we can only make predictions about how the dust attenuation behaves farther into the history of the Universe, at redshift $z \sim 10$. We give Eq. \eqref{eq:Afuv_z_M*} as a recipe for predicting the dust attenuation of galaxies given their redshift and stellar mass. This can be further used to give an estimate of the dust attenuation where no data is available, as well as to make predictions and simulations to further push the limits of the knowledge of this field.

\section{Conclusions}

In this work, we estimate the evolution with redshift of the dust attenuation in the FUV by first exploring the evolution of the relationship between the dust attenuation and the stellar mass throughout cosmic times. The evolution of the $A_{FUV}-M_*$ relationship has been debated in the literature. However this paper strongly suggest that we need to assume that the $A_{FUV}-M_*$ relationship does evolve with redshift, and we base our further studies upon this hypothesis.

An additional interesting point is that predictions can be made using the prescriptions presented in this paper. These predictions can be tested using data and the \textit{JWST} and new deep sub-millimetre and millimetre facilities in a relatively near future.

We can summarise this work with the following conclusions: 
\begin{enumerate}
\item 
The $A_{FUV}-M_*$ relationship needs to be described with a more complicated function as opposed to the consensus linear (in terms of $\log M_*$) relationship, such as the one proposed in Eq. \eqref{eq:Afuv_M*_break}. Such a function needs to be able to incorporate the influence of the low mass galaxies on the global average of the dust attenuation.

\item
Assuming the $A_{FUV}-M_*$ relationship does not evolve with redshift is not consistent with other studies concerning the evolution of the dust attenuation. On the other hand, starting from the assumption that this relation is not the same at all cosmic times gives results similar to the ones found by groups studying the same phenomenon by the use of different methods. 

\item

The $A_{FUV}-M_*$ relationship for lower stellar masses has a large scatter, with an average value which is likely to be larger than zero throughout most of the cosmic times. The physical origin of this offset cannot be derived from the present data. However, some works listed in Sect.~\ref{sect:discussion} suggest that this flattening can have different origins that we need to explore: simply a large dust content in these low-mass galaxies but this seems unlikely, the stars-dust geometry, the dust-to-metal ratio.

\end{enumerate}

\section*{Acknowledgements}

The authors would like to thank Xiangcheng Ma, Christopher Hayward, and Ambra Nanni for their helpful comments and discussion. JB is also grateful for the advice and suggestions given by Junais. Finally, we thank the anonymous referee for the constructive criticism that helped bring out the best of our work.



\bibliographystyle{mnras} 
\bibliography{AfuvMstarZ.bib} 

\begin{appendix}
\section{Computation of the average dust attenuation}\label{sect:appendix_MF}

 \begin{figure}
   \centering
    \resizebox{0.8\hsize}{!}{\includegraphics[width=17cm]{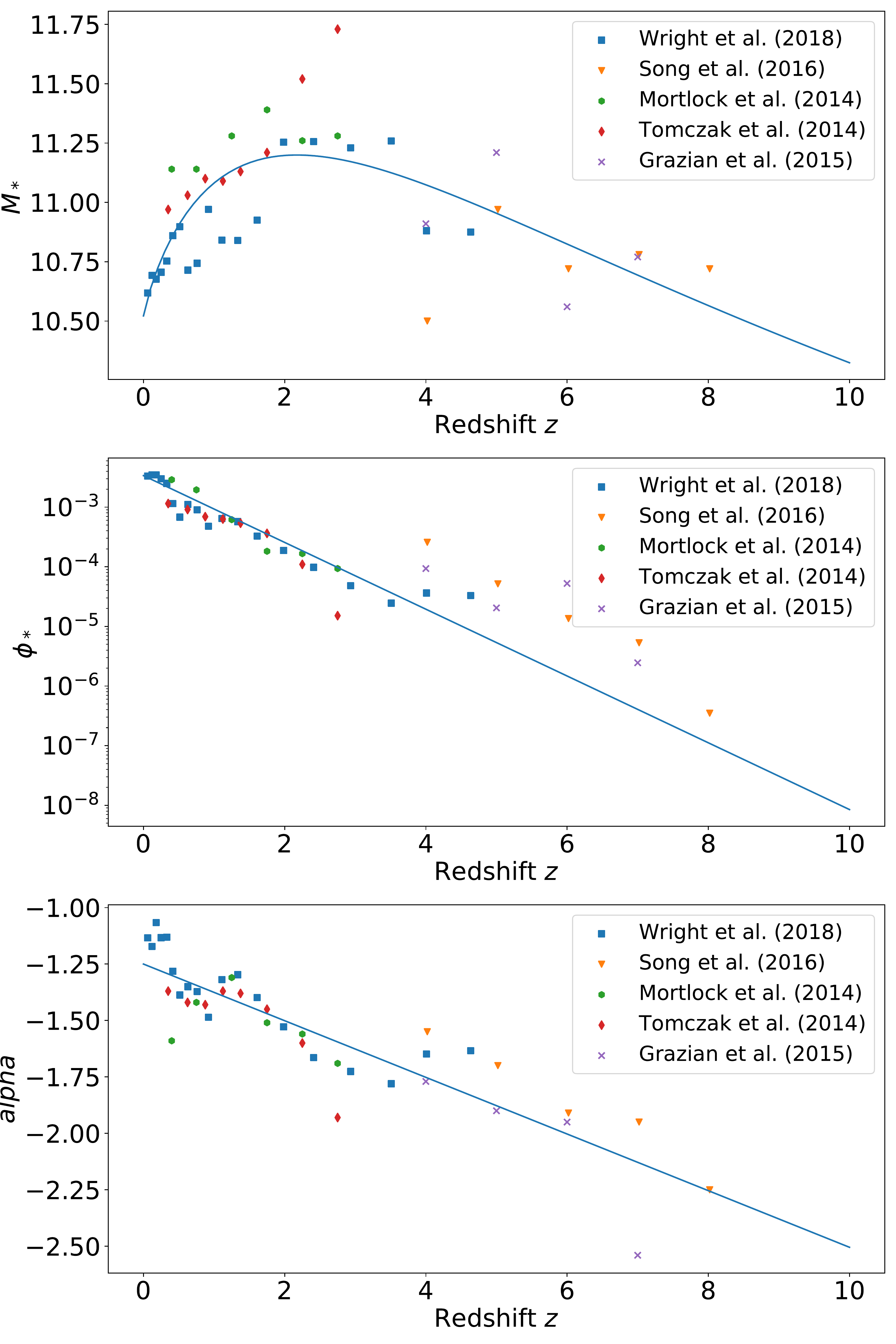}}
     \caption{The fitting of the Schechter parameters. The functions used are $\mathcal{M^*} = ({k_1} + {k_2}z) / (1 + (z/{k_3})^{k_4})$, where $k_1 = 10.52$, $k_2 = 2.38$, $k_3 = 4.80$ and $k_4 = 1.15$; $\log \phi_* = {(l_1-0.56z)}$, with $l_1 = -2.47$; $\alpha = {m_1} + {m_2}z$, with parameters $m_1 = -1.25$ and $m_2 = -0.13$.}
     \label{fig:schechter}
\end{figure}

Fig. \ref{fig:a_fit} indirectly provides an information on  the evolution of the dust attenuation with redshift. This makes it difficult to compare to the literature. We proceed by computing the average dust attenuation for all of the stellar masses, and in this section we present the recipe that we followed to do that. 

By definition, if the dust attenuation of a galaxy is a function of its stellar mass $A_{FUV} (M_*)$, then the mean of this function would be: 

\begin{align}\label{eq:mean_value}
\overline{A_{FUV}} = 
\frac{
 \int_{M*_\mathrm{min}}^{M*_\mathrm{max}} A_{FUV} (M_*) \phi(M_*) \mathrm{d}M_*
}{
 \int_{M*_\mathrm{min}}^{M*_\mathrm{max}} \phi(M_*) \mathrm{d}M_*
}
\end{align}

Here, $\phi(M_*)$ is the mass function (MF), which in this case acts as a normalisation. We use the functional form  of \citet{schechter_analytic_1976}, with $\phi_*, \mathcal{M^*}$ and $\alpha$ the Schechter parameters:

\begin{align}\label{eq:MF_used}
\phi \mathrm{d}M = 
\phi^* 10 ^ { (1 + \alpha) (\log M_* - \log \mathcal{M^*}) } 
\exp \Big[  -10 ^ {(\log M_* - \log \mathcal{M^*})}  \Big]  \mathrm{d}M
\end{align}

The evolution of the dust attenuation-stellar mass relationship can now be expressed through the evolution of the average dust attenuation with redshift. Studying the evolution of the average dust attenuation requires an estimation of the value of Eq. \eqref{eq:mean_value} for each redshift. This, in turn, requires the knowledge of the evolution of the MFs, for which we used the MFs of \cite{tomczak_galaxy_2014}, \cite{grazian_galaxy_2015},  \cite{mortlock_deconstructing_2015}, \cite{song_evolution_2016}, and \cite{wright_gama/g10-cosmos/3d-hst_2018}. We fit the values of the \cite{schechter_analytic_1976} parameters given in these papers to be able to retrieve their value at any given redshift (Fig. \ref{fig:schechter}); the values for the $\mathcal{M^*}$ parameter are fitted with the function $\mathcal{M^*} = ({k_1} + {k_2}z) / (1 + (z/{k_3})^{k_4})$, and the best-fitting is for $k_1 = 10.52$, $k_2 = 2.38$, $k_3 = 4.80$ and $k_4 = 1.15$, for $\phi_*$ we have $\log\phi_* = {(l_1-0.56z)}$, with $l_1 = -2.47$ giving the best-fitting, and for $\alpha$ the function is a line $\alpha = {m_1} + {m_2}z$, with parameters $m_1 = -1.25$ and $m_2 = -0.13$.

We set up a grid of redshifts, and for each value $z_i$ we calculate the dust attenuation using the model we have chosen for the $\overline{A_{FUV}}-z$ relationship, with the corresponding value of the coefficient $a(z_i)$. For the same $z_i$ we estimate the \cite{schechter_analytic_1976} parameters, and compute the corresponding MF. We then use the MF as the weight for calculating the average dust attenuation $\overline{A_{FUV}}$, according to Eq. \eqref{eq:mean_value}. 
The results of {this computation} are represented in Sect. \ref{sect:A_FUV_z}.

  \section{Dust attenuation as a function of both redshift and stellar mass} \label{sect:appendix_3d}

The work presented in this paper strives to combine the dependence of the dust attenuation on stellar mass and its evolution with redshift. The result of this unification is a three-dimensional model for the dust attenuation as a function of both stellar mass and redshift, $A_{{FUV}}(z, M_*)$, that is a surface in a 3D space. The stellar mass and the redshift are independent variables, while the dust attenuation depends on both of these values. This gives us the ability to estimate the dust attenuation of any galaxy from knowing its stellar mass and redshift. 

We already have a functional form of both the dependencies we require, $A_{{FUV}}(M_*)$ and $A_{{FUV}}(z)$ by fitting the parameter $a(z)$ and we can directly replace it in Eq. \eqref{eq:Afuv_M*}. Thus, we get the relation for $A_{{FUV}}(z, M_*)$ if we put together Eqs. \eqref{eq:Afuv_M*} and \eqref{eq:Afuv_z} to get: 

\begin{align}\label{eq:Afuv_z_M*}
A_{{FUV}} = & \left(z+\gamma\right)\cdot \alpha^{\left(\beta-\left(z+\gamma\right)\right)} \nonumber \\ &\times 
\begin{cases}
1.1 , &\log M_* < 9.8\\ 
\log M_* - 8.7 , & \log M_* > 9.8
\end{cases}
\end{align}

\begin{figure}
 \resizebox{\hsize}{!}{\includegraphics{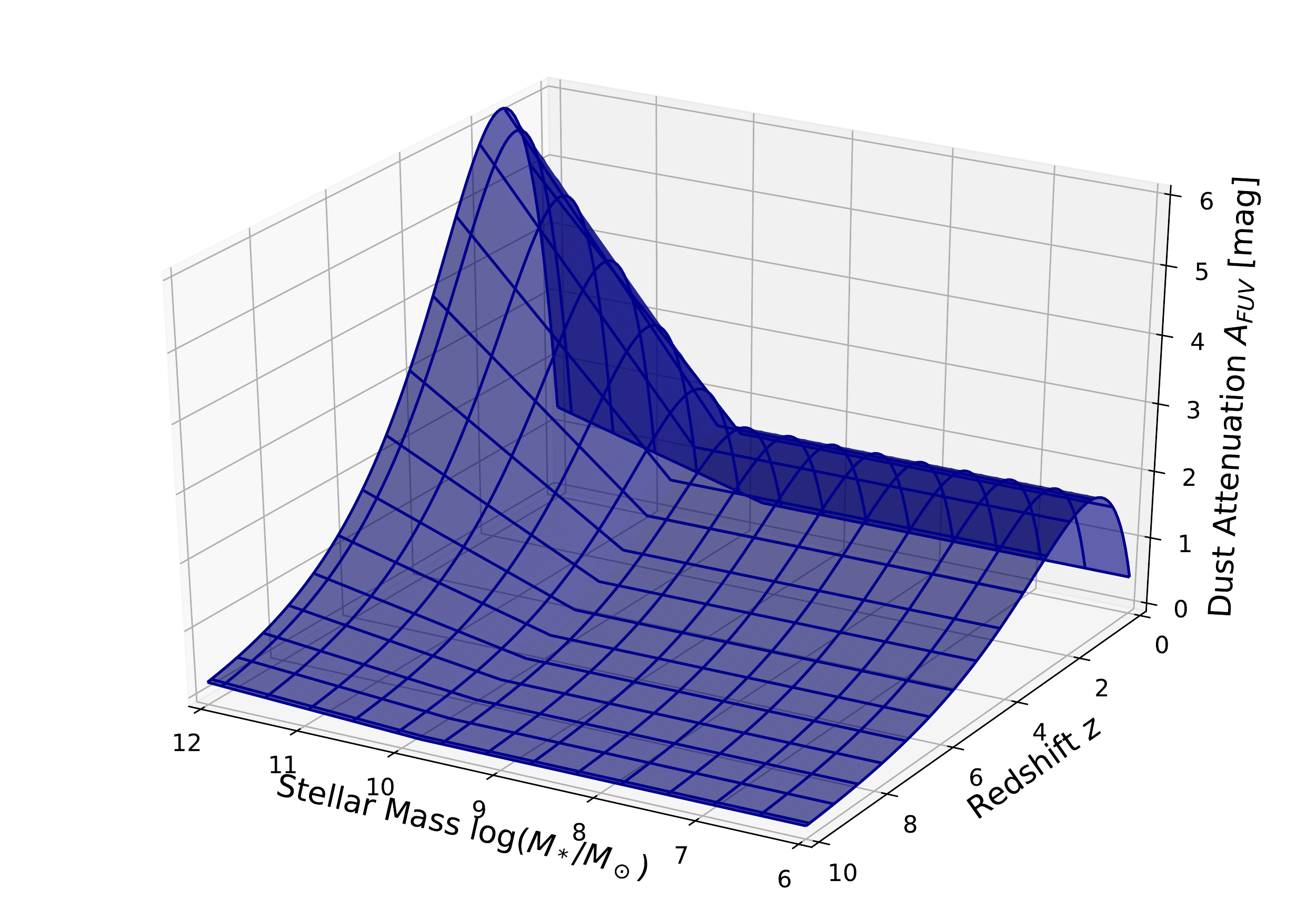}}
  \caption{The dependence of the dust attenuation in the UV on stellar mass and redshift. The surface represents the model shown in Eq. \eqref{eq:Afuv_z_M*}. If we take, for example, any value $\log M_* =$ const., we retrieve the dependence given by Eq. \eqref{eq:Afuv_z}, shown in Fig. \ref{fig:a_fit}. Similarly, for any value of the redshift, we retrieve the models of Eq. \eqref{eq:Afuv_M*_break},  shown in Fig. \ref{fig:AFUV_Mstar_data}. }
 \label{fig:3d}
\end{figure}

The parameters of this function are the same ones that are determined with the models discussed in Sect. \ref{sect:modify_Afuv_M}, and thus, their values remain the same. The 3D plot of this relation is shown in Fig. \ref{fig:3d}.

\end{appendix}

\label{lastpage}

\end{document}